\documentclass[aps,prb,showpacs,twocolumn,groupedaddress,floatfix]{revtex4}
\usepackage{graphicx}
\usepackage{amssymb,amsmath}
\usepackage{ulem}

\begin{document}

\title{Ideal Strength of Doped Graphene}
\author{S. J. Woo}
\author{Young-Woo Son}
\email{hand@kias.re.kr}
\affiliation{Korea Institute for Advanced Study, Seoul 130-722, Korea}

\date{\today}
\begin{abstract}
While the mechanical distortions change the electronic properties
of graphene significantly, the effects of electronic manipulation
on its mechanical properties have not been known.
Using first-principles calculation methods,
we show that,
when graphene expands isotropically under equibiaxial strain,
both the electron and hole doping can maintain or
improve its ideal strength slightly
and enhance the critical breaking strain dramatically.
Contrary to the isotropic expansions,
the electron doping decreases the ideal strength as well as critical strain 
of uniaxially strained graphene
while the hole doping increases both.
Distinct failure mechanisms depending on type of strains
are shown to be origins of the different doping induced mechanical stabilities.
Our findings may resolve a contradiction 
between recent experimental and theoretical results
on the strength of graphene.
\end{abstract}

\maketitle
\section{Introduction}
A defect-free infinite crystal becomes mechanically unstable at a stress with
a corresponding critical strain. Such a stress
is the ideal strength of a material which is an inherent property
of a given atomic and electronic structure
as well as a natural upper bound on its strength~\cite{kelly,morris}.
While the variation of electron density can alter ideal strength of some materials~\cite{jhi},
those manipulations usually involve stoichiometric change
which may alter the nature of chemical bonding between atoms in crystal.
In this regard, graphene, a two-dimensional crystal with atomic thickness,
is an ideal material to investigate electronic control of ideal strength because
it allows very high level of charge doping without any sacrifice of atomic integrity
through either field-effect transistor setups~\cite{Novoselov,PKim} or
alkali metal depositions~\cite{Eli}.
Although there have been many studies about effects of mechanical perturbation on
electron physics in graphene~\cite{Guinea,Pereira,Peres,SMChoi,Vozmediano,Levy},
the effect of electron density variation on the mechanical properties of graphene
has not been known yet.

Graphene is also known to be the strongest two dimensional (2D) material~\cite{CLee}
and can maintain its strength 
in the presence of grain boundaries~\cite{ruoff}.
A recent nanoindentation experiment~\cite{CLee} reports
that the intrinsic strength of graphene is 42 N m$^{-1}$ at the
nominal equibiaxial breaking strain of 0.225.
This was followed by a theoretical calculation~\cite{CAMarianetti} 
showing the quite smaller breaking strain of 0.15
than the measured value.
It is unusual since materials under strain typically fail
before they reach the ideal strength due to various reasons.
Therefore considerations on the other external effects such as
a doping are called for to understand the ideal strength of graphene further
since typical graphene samples in experiments
are doped.

In this work, we report a theoretical study showing
that graphene under equibiaxial strain becomes
to be stronger as doping increases with both electron
and hole doping.
Though the ideal strength under equibiaxial strain increases slightly
by $1\sim6~\%$ depending on doping type and amounts, the corresponding
critical breaking strain increases dramatically by $\sim19~\%$
within the doping level of $\sim1.1\times 10^{14} /{\rm cm}^2$ which
is accessible in experiments~\cite{PKim,Eli}.
We show that the failure of isotropically expanded 
graphene through the $A'_1$ phonon mode softening~\cite{CAMarianetti} 
or the two-dimensional Peierls instability~\cite{SHLee}
is overcome by doping because the Fermi level ($E_F$) of doped
graphene lies outside the energy gap associated with the $A'_1$ phonon.
On the contrary, under uniaxial strain
invoking elastic failure of graphene~\cite{CAMarianetti},
we find that electron doping weakens graphene by $5\sim7\%$
while hole doping strengthens one by $3\sim6\%$
and that the corresponding breaking strains also change a lot by $37\%$ 
within similar doping levels for the equibiaxial strain cases.
Because the elastic failure under uniaxial strain is associated with
occupation of the $\sigma^*$ band~\cite{SMChoi},
the asymmetric dependence of ideal strength on electron and hole doping
can be explained in term of doping induced occupation (electron doping) or
depopulaton (hole doping) of the $\sigma^*$ band of uniaxially strained graphene.

This paper is organized as follows. In Sec.~\ref{sec:method}, we introduce calculation
methods used in this work. After introducing models for the first-principles calculations, 
we examine the ideal strength of graphene under equibiaxial strains with various
electron and hole dopings in Sec.~\ref{sec:equi}. Then, we also study the ideal
strength of doped graphene under uniaxial strain with different strain directions
and reveal origins of variations in ideal strength depends on strain as well as dopings
in Sec.~\ref{sec:uni}. In Sec.~\ref{sec:dis}, we discuss a possible resolution
on discrepancy between experiment~\cite{CLee} and theoretical works~\cite{CAMarianetti,SHLee}
and conclude this work.

\section{Calculation Methods\label{sec:method}}
Our first-principles calculations~\cite{PGiannozzi} are carried out with plane wave basis
and norm-conserving pseudopotentials~\cite{NTroullier}.
The local density approximation (LDA) is used for the exchange-correlation functional~\cite{JPPerdew}.
Phonon spectrum is calculated using density functional perturbation theory~\cite{Baroni,PGiannozzi}
with $12\times12\times1$ $q$-point sampling.
The energy cutoff for the basis set expansion is $80$~Ry.
The $k$-point grid of $48\times48\times1$ is used.
Doping is simulated within the rigid band approximation
which is successfully used in recent studies on Kohn anomaly of doped carbon
nanotubes~\cite{Bohnen} and graphane~\cite{Savini} respectively.
Near the Kohn anomaly where the phonon frequency changes very abruptly
in a narrow momentum range, the explicit phonon calculations are additionally done
to avoid the inherent errors in interpolated phonon dispersions.

\section{Ideal Strength of doped Graphene under equibiaxial strain\label{sec:equi}}
\begin{figure}[t]
 \includegraphics[width=\columnwidth]{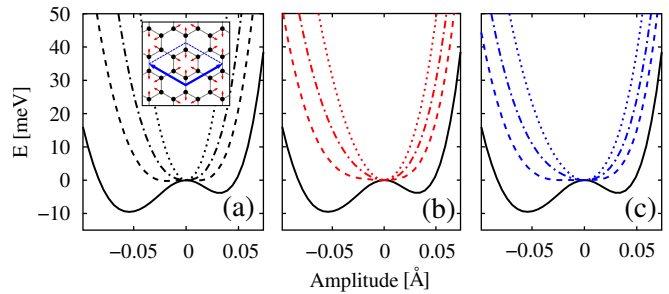}
 \caption{(Color online)
 (a) The total energy variation as a function of $A'_1$ phonon amplitude under
equibiaxial strains of $\epsilon=$ 0.12, 0.14, 0.15, and 0.16
from top to bottom lines.
Inset shows schematic atomic displacements (red arrows) 
of the $A'_1$ phonon mode and the $K$-point supercell (blue arrows).
(b) and (c) show the same curves as in (a) at a fixed strain, $\epsilon=0.16$,
with changing electron and hole doping from top to bottom lines,
$n_e (n_h)=0.04$, $0.02$, $0.01$, $0.0$ respectively.
 }
  \label{e_vs_a1amp_strain_doping}
\end{figure}

When graphene expands isotropically, the $A'_1$ phonon mode
at the $K$ point softens and then becomes to be unstable 
at a certain critical equibiaxial strain~\cite{CAMarianetti}.
The inset of Fig.~\ref{e_vs_a1amp_strain_doping}~(a) show schematically
how the crystal structure of graphene is deformed when the $A'_1$ phonon is excited.
Here, the strain is defined by $\epsilon = (a-a_0)/a_0$, where $a$ and $a_0$ are
the lattice constant of equibiaxially strained graphene and that of pristine graphene respectively.
In Fig.~\ref{e_vs_a1amp_strain_doping}~(a), the total energy variations
as a function of the amplitude of $A'_1$ phonon mode
show that, for $\epsilon>0.15$, the hexagonally deformed structure
is energetically more stable than the symmetric one~\cite{CAMarianetti}.
For these caculations we use the $K$-point supercell
in which six carbon atoms are included [inset of Fig. 1(a)]~\cite{CAMarianetti}.
This can be understood as two-dimensional extension of Peierls instability~\cite{SHLee}.
Undoped graphene has its $E_F$ at the Dirac point and
the excitaion of $A'_1$ phonon mode generates a dynamical energy gap
at the point~\cite{SHLee,Samsonidze} realizing
the Kohn anomaly~\cite{Samsonidze,Piscanec}.
As graphene expands over critical equibiaxial strain,
the energy gain through formation of static energy gap at the $K$-point
overcomes the energy cost for ion-ion interaction
so that the hexagonal network of graphene breakdowns~\cite{CAMarianetti,SHLee}.
Therefore, if the $E_F$ shifts away from the Dirac point by doping,
the Peierls instability will be avoided by hardening the $A'_1$ phonon so that graphene
becomes more resistive against the applied equibiaxial strain.

To confirm this hypotehsis,
we calculate the total energy of doped graphene
when the $A'_1$ phonon mode is excited.
As discussed, Fig.~\ref{e_vs_a1amp_strain_doping}~(a) shows 
the lattice instability of undoped graphene over a critical 
equibiaxial strain ($\epsilon > 0.15$).
The same calculations were performed for doped graphene 
at various doping levels with a fixed overcritical
strain of $\epsilon = 0.16$ [Figs. \ref{e_vs_a1amp_strain_doping}~(b) and (c)].
The electron (hole) doping level, $n_{e} (n_h)$, is defined by the number of additional
electrons (holes) per a carbon atom.
We note that $n_{e}=0.01$ corresponds to $3.848\times10^{13}$ electron$/$cm$^2$.
As shown in Figs.~\ref{e_vs_a1amp_strain_doping}~(b) and (c),
the electron and hole dopings indeed recover the stability of $A'_1$
phonon mode even at the strain of $\epsilon = 0.16$ under which pristine graphene
is already broken.

\begin{figure}[t]
 \includegraphics[width=\columnwidth]{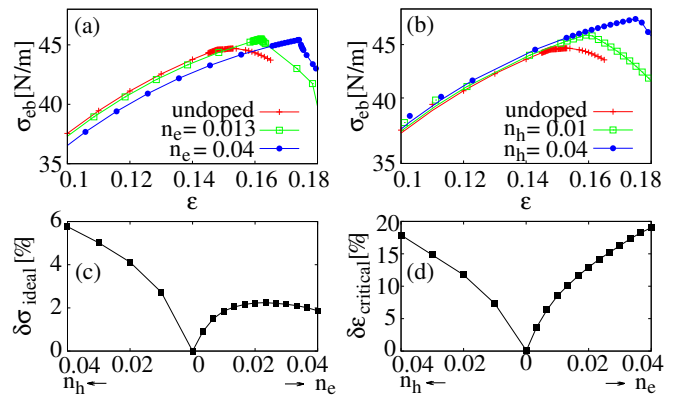}
 \caption{(Color online)
(a) and (b) show the stress-strain relationship of graphene under equibiaxial strains 
	with different electron and hole doping levels respectively.
The equibiaxial strain is defined by
$\sigma_{\rm eb}\equiv\sqrt{\sigma_x^2+\sigma_y^2}/\sqrt{2}$.
The variations (in percentage) of (c) ideal strength ($\delta\sigma_{\rm ideal}$) and
(d) corresponding critical breaking strain ($\delta\epsilon_{\rm critical}$)
of doped graphene with respect to those of undoped one 
as a function of doping.
 }
  \label{stress_strain}
\end{figure}

To obtain complete ideal strengths ($\sigma_{\rm ideal}$) 
of doped graphene, the stress-strain relationships are needed,
in which the maximum stress at the critical strain ($\epsilon_{\rm critical}$)
is the $\sigma_{\rm ideal}$ of doped graphene.
In Figs. \ref{stress_strain}~(a) and (b), 
stress maximum in
the stress-strain relationship of doped graphene shifts
toward larger strain as the doping increases independent of the doping type.
The $\sigma_{\rm ideal}$ of undoped graphene 
is 44.6 N/m agreeing well with
recent studies~\cite{CLee,CAMarianetti}.
The critical breaking strain of 0.148, however, is much smaller than
the measured value~\cite{CLee,CAMarianetti, SHLee}.
As $n_e$ increases from 0.00 to 0.04, 
$\sigma_{\rm ideal}$ increases by $\sim2\%$ ($n_e=0.02$)
and then decreases. 
As $n_h$ increases from 0.00 to 0.04, $\sigma_{\rm ideal}$ 
increases monotonically by $\sim6\%$ [Fig. 2(c)].
Contrary to that, the $\epsilon_{\rm critical}$
shows a dramatic increase by $\sim19\%$ ($\epsilon_{\rm critical}\sim 0.19$)
as both $n_e$ and $n_h$ increase to 0.04 [Fig. 2(d)].

\begin{figure}[t]
 \includegraphics[width=\columnwidth]{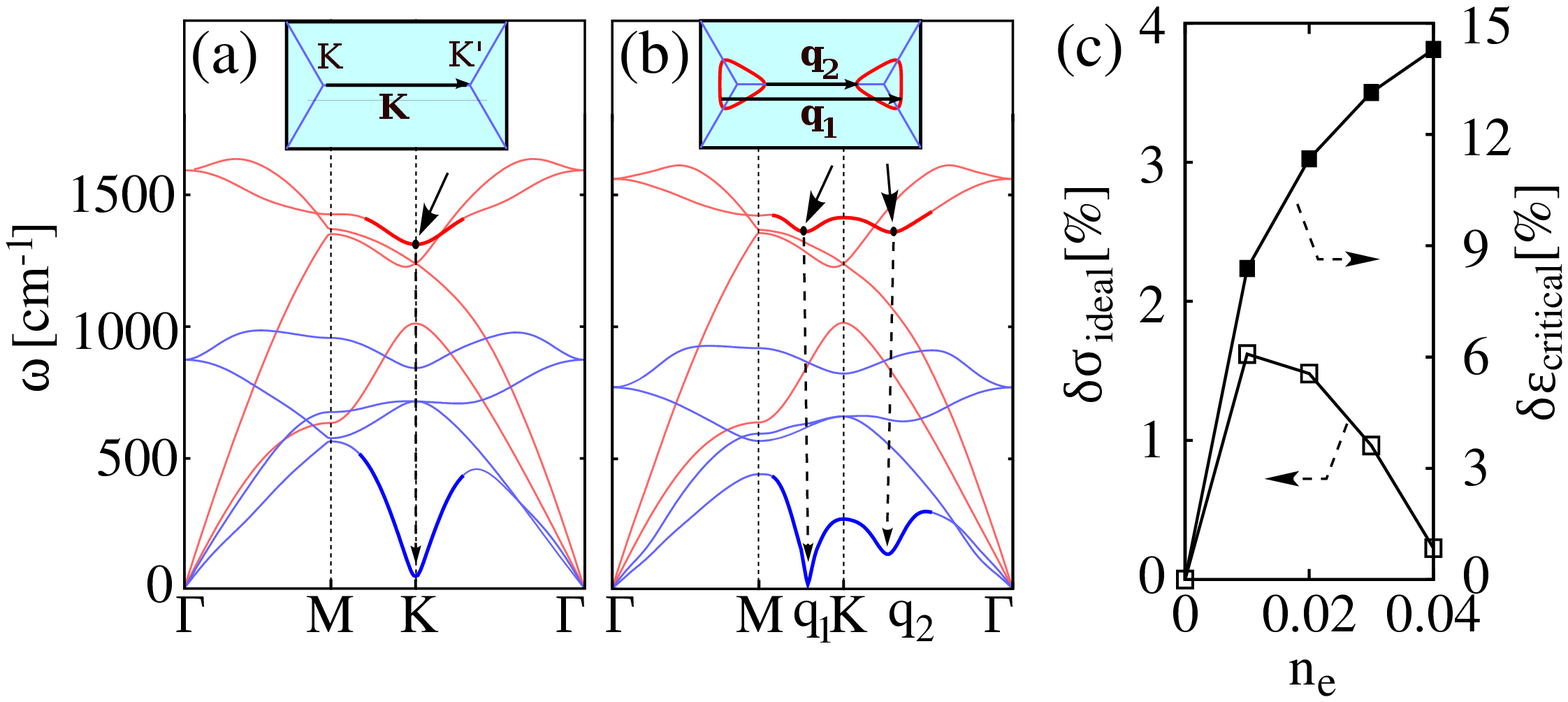}
 \caption{(Color online)
(a) Phonon dispersion of undoped graphene with equibiaxial strain $\epsilon=0$ (red lines)
	and 0.148 (blue). 
(b) Same as in (a) except for doped graphene ($n_e=0.04$) with $\epsilon=0$ (red) and 
0.167 (blue).
In both panels, the black arrows indicate the Kohn anomaly and 
the insets show the vectors (black arrows) connecting Fermi surfaces (red lines) 
related with the Kohn anomaly.
Here we do not show irrelevant out-of-plane phonon mode spectrum.
(c) $\delta\sigma_{\rm ideal}$ 
and $\delta\epsilon_{\rm critical}$ (left and right ordiates) 
obtained from phonon dispersions as function of electron doping $n_e$ (abscissa).
}
  \label{phonon_dispersion}
\end{figure}

We have further calculated phonon dispersions in order to
see how the phonon softening at the $K$-point is affected by the doping.
Softening of phonon to the negative phonon frequency
may precede or indicate the crystal instability~\cite{Chatterbuck}.
Fig. \ref{phonon_dispersion}~(a) shows the phonon dispersion
of graphene with $\epsilon=0$ (red) and $\epsilon=0.148$ (blue)
agreeing well with a previous study~\cite{CAMarianetti}.
The frequency dip at the $K$-point associated with the $A'_1$ mode
is the Kohn anomaly~\cite{Samsonidze,Piscanec}.
We note that the LDA is not quantitatively perfect
to describe the anomaly~\cite{Lazzeri} but is sufficient to describe
the softening within certain errors~\cite{CAMarianetti,Bohnen,Savini}.
The Kohn anomaly may occur for a phonon of momenta $\bf{q}$ 
when two electronic states of momentum
${\bf k} _1$ and ${\bf k} _1+\bf{q}$ are on the Fermi surface~\cite{WKohn,Piscanec}.
In undoped graphene, the Fermi surface is at the Dirac points located at $\bf K$ and
${\bf K}'=2\bf K$ so that ${\bf q}=0$ ($\Gamma$-point) and ${\bf q}=\bf{K}$ ($K$-point)
are allowed for the anomaly.
The inset in Fig.~\ref{phonon_dispersion}~(a)
shows the corresponding vector $\bf K$ in the BZ.
As graphene doped, the momentum at which the anomaly occurs
shifts from the $K$-point and splitts into two points.
The $A'_1$ phonon hardens also because of change of the $E_F$.
In electron-doped graphene ($n_e =0.04$),
the Kohn anomaly occurs both at $q_1$
between $M$ and $K$ and at $q_2$ between $K$ and $\Gamma$ [Fig.~\ref{phonon_dispersion}~(b)].
The inset in Fig.~\ref{phonon_dispersion}~(b) shows the correspoding
vectors ${\bf q}_1$ and ${\bf q}_2$ connecting the two electronic states
at Fermi surfaces of two valleys in doped graphene.
The phonon at $q_1$ softens to zero frequency 
at the critical equibiaxial strain of $\epsilon=0.167$ [Fig.~\ref{phonon_dispersion}~(b)].
This indicates that the doped graphene becomes structurally unstable
through the less symmetric phonon mode at $q_1$ than $A'_1$ mode.
The strain value of 0.167 for $n_e=0.04$ doping
differs from the corresponding critical strain of 0.179
obtained by the stress-strain relationship [Fig. \ref{stress_strain} (d)].

By taking stress and strain values when the softened phonon mode touches zero frequency,
we can also obtain the doping dependent ideal strengths and the corresponding critical strains.
As shown in Fig.~\ref{phonon_dispersion} (c), the overall behaviors of $\delta\sigma_{\rm ideal}$
and $\delta\epsilon_{\rm critical}$ are similar to those obtained (Figs. 2 (c) and (d))
from the stress-strain curves. However, the ideal strength increases only by $\sim1.7\%$ and then
decreases back to the original value as $n_e$ reaches 0.04 [Fig. 3(c)]
due to phonon softening at $q_1$.
The corresponding breaking strain shows the significant increase by $\sim15\%$.
So, we can conclude that under equibiaxial strain the electron and hole doping
improve the ideal strength slightly or at least maintain
its undoped value while the corresponding critical strain increases dramatically by doping.
These interesting behaviors of stress and strain of doped graphene can be understood by noting that
the ideal strength is mostly determined by $\sigma$ bonds between carbon atoms that does not be
affected much by the doping
but that the breaking point of strain is determined by $\pi$ bands owing to the Kohn anomaly
that changes greatly according to the doping.

\begin{figure}[t]
 \includegraphics[width=\columnwidth]{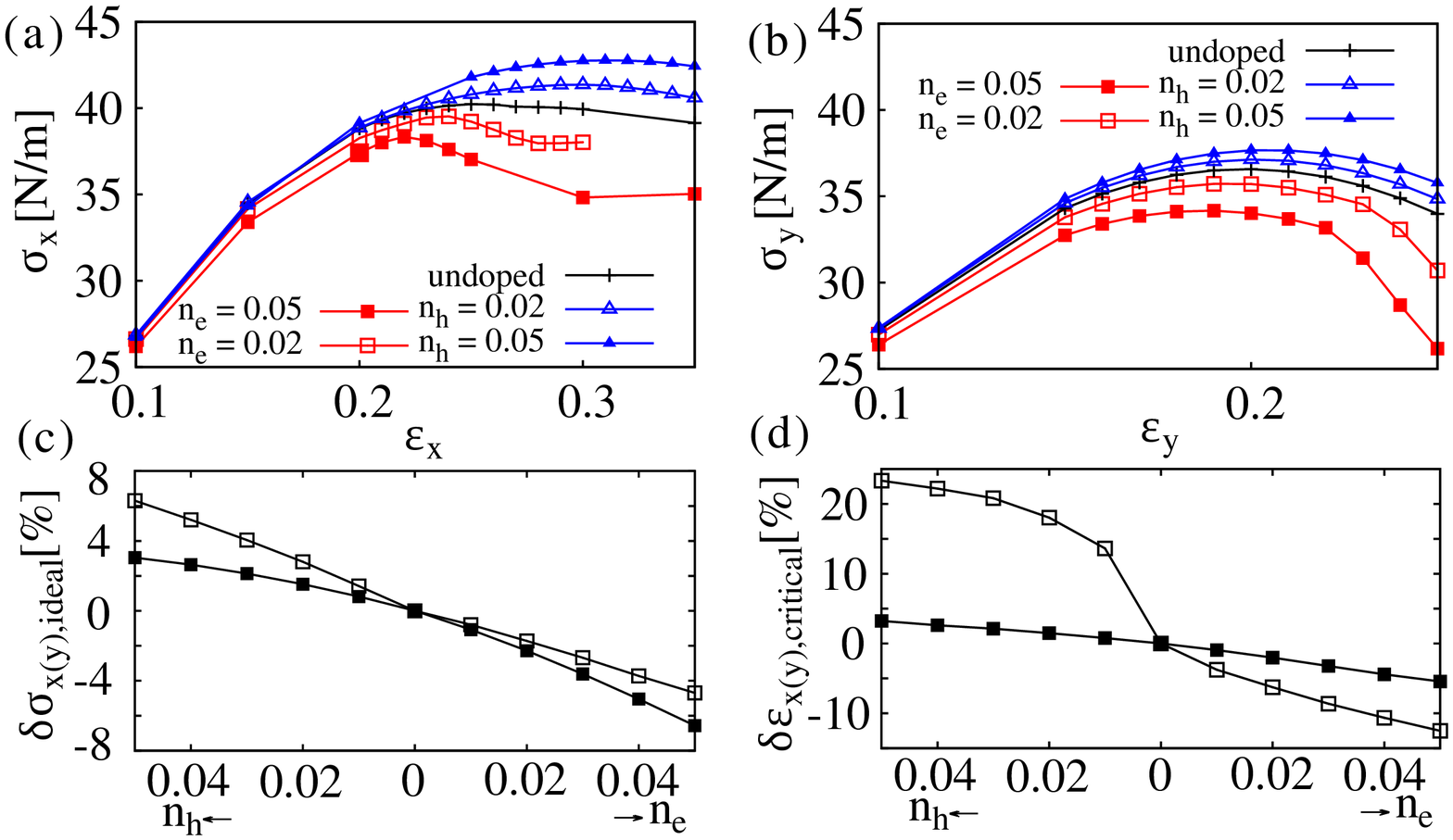}
 \caption{(Color online)
 Stress-strain curves for graphene under (a) zigzag ($\epsilon_x$) and
(b) armchair ($\epsilon_y$) strain with various doping levels.
The variations (in percentage) of (c) ideal strength and (d) critical breaking strain as a function
of the doping level for graphene under the $\epsilon_x$ (open rectangles) and $\epsilon_y$ (solid ones)
respectively.
 }
  \label{stress_strain_uniaxial}
\end{figure}

\section{Ideal Strength of doped graphene under uniaxial strain\label{sec:uni}}
We have analyzed the ideal strength of doped graphene under uniaxial strains.
Unlike the equibiaxial strain cases, 
the uniaxial strain breaks the crystal symmetry of graphene so that
Dirac cone moves away from $K$-point
and distorts into anisotropic form~\cite{Peres,SMChoi}.
Here we only consider uniaxial strains either
along zigzag or armchair direction~\cite{Peres,SMChoi}
(called as zigzag ($\epsilon_x$) and armchair ($\epsilon_y$) strain respectively)
and calculate the stress-strain relationship to find out the doping dependent ideal
strength.
For these calculations, the primitive unit cell including two carbon atoms is used
since uniaxially strained graphene breaks down
elastically unlike equibiaxial strain cases~\cite{CAMarianetti}.
Fig. \ref{stress_strain_uniaxial} (a) and (b) show the stress-strain curves
for the zigzag and armchair strain respectively with different doping levels.
It is found that the ideal strength under zigzag strain
is more sensitive on the doping than that under the armchair one.
Figs. ~\ref{stress_strain_uniaxial} (c) and (d) show the change of
ideal strength and critical strain as function of dopings respectively.
Contrary to the equibiaxial strain case,
the response of graphene under uniaxial strain
is asymmetric with respect to the electron and hole doping:
graphene becomes stronger with the hole doping while it becomes weaker
under the electron doping.
For undoped graphene, the maximum stress of $\sigma_{\rm ideal}=40.2$ (36.6) occurs
at $\epsilon_{\rm critical}=0.253$ (0.198) for the zigzag (armchair strain)
agreeing with a previous study~\cite{Liu}.
Similar to the equibiaxial strain case,
the doping induce a huge change of the strain values but not the ideal strength
[Figs. 4(c) and (d)].
Under zigzag strain, the critical breaking strain increases
from 0.253 to 0.311 ($23\%$ increase) as $n_h$ reaches 0.05,
and decreases to 0.22 ($13\%$ decrease) as $n_e$ reaches 0.05.
On the other hand, under the armchair strain,
it increases only by $3.5\%$ and decreases by $5\%$ in the same condition.

\begin{figure}[t]
 \includegraphics[width=\columnwidth]{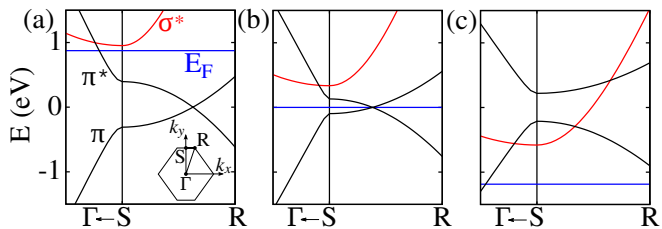}
 \caption{(Color onlie)
	 Electronic band structures of graphene under
(a) the zigzag strain of $\epsilon_x = 0.22$ with $n_e=0.05$,
(b) $\epsilon_x = 0.25$ without doping, and
(c) $\epsilon_x = 0.30$ with $n_h=0.05$.
The red and black lines are the $\sigma^*$ and $\pi (\pi^*)$ bands respectively
and the blue is the Fermi level.
The inset in (a) shows the high symmetry points
of BZ of graphene with $\epsilon_x$.
 }
  \label{eband}
\end{figure}

Especially under the zigzag strain, the two Dirac cones 
at the $K$- and $K'$-point approach each other and eventually merge
after a critical strain opening a small energy gap 
at the merged point~\cite{Peres,SMChoi,Goerbig}.
At the same time, however, the energy of $\sigma^*$ band
is lowered toward the $E_F$~\cite{SMChoi} signaling the breakdown
of crystal structure.
Our calculation shows that the stress maximum occurs
as the unoccoupied $\sigma^*$ band reaches the $E_F$.
Figs. \ref{eband} (a), (b) and (c) show
the electronic band structure for electron-doped, undoped,
and hole-doped graphene under the zigzag strain respectively.
For the electron doping with $n_e=0.05$ the $\sigma^*$ band is closer
to the $E_F$ compared with one in undoped graphene under zigzag strain
so that the band is filled up at a relatively lower strain of
$\epsilon_x = 0.22$ [Fig. \ref{eband} (b)].
Contrary to that, for the hole doping with $n_h=0.05$ the $\sigma^*$ band moves away
from the $E_F$ so that it requies relatively
higher strain of $\epsilon_x = 0.30$ for the band to be filled up [Fig. \ref{eband}~(c)].
This explains the physical origin of electron-hole doping asymmetry of
the ideal strength of doped graphene under zigzag strain.

\section{Discussion and conclusion\label{sec:dis}}
We discuss a possible resolution of the discrepancy between
the recent experiment~\cite{CLee} and theories~\cite{CAMarianetti, SHLee} on the 
strength of graphene. 
From our calculations, the doped graphene
can exhibit a theoretical ideal strength ($>$ 40 N m$^{-1}$)
with a enhanced critical equibiaxial strain.
Though the amount of doping in the experiment~\cite{CLee} 
has not been known, the combination of several unknown experiment factors
such as anisotropy in applied forces~\cite{CLee},
formation of dislocation~\cite{ruoff}, and non-uniformity in number of layers
as well as doping may enhance the critical strain value further than
the undoped maximum of $\epsilon=0.148$.

In conclusion, we show that
the doping induces significant strain-dependent variations in the mechanical stability of graphene.
Thus, our calculations set new bounds on the ideal strength and critical strain
of graphene in realistic situations.
We believe that our study not only highlights interesting
interplay between electronic and mechanical properties of graphene
but also will be useful to explore such properties 
in other 2D crystals~\cite{Nobel}.

{\it Note added.} After submission, we became aware of a recent paper~\cite{ChenSi}
having an overlap with a part of our work which was also reported recently~\cite{APS}.

\begin{acknowledgments}
This work is supported by the NRF of Korea grant funded by MEST
(QMMRC, No. R11-2008-053-01002-0 and CASE, No. 2011-0031640).
Computations are supported by
KISTI Supercomputing Center (Project No. KSC-2011-C1-21)
and the CAC of KIAS.
\end{acknowledgments}

\end{document}